\documentclass[10pt, conference, compsocconf]{IEEEtran}
\IEEEoverridecommandlockouts

\def\IEEEbibitemsep{0pt plus .5pt}

\usepackage{cite}

\ifCLASSINFOpdf
  \usepackage[pdftex]{graphicx}
\else
\fi

\usepackage[cmex10]{amsmath}

\usepackage{algorithmic}

\usepackage{tikz}
\usepackage[ruled, linesnumbered]{algorithm2e}
\usepackage{subcaption}
\usepackage{multirow}
\usepackage{cleveref}

\newcommand\copyrighttext{%
  \footnotesize \textcopyright 2021 IEEE. Personal use of this material is permitted.
  Permission from IEEE must be obtained for all other uses, in any current or future
  media, including reprinting/republishing this material for advertising or promotional
  purposes, creating new collective works, for resale or redistribution to servers or
  lists, or reuse of any copyrighted component of this work in other works.
}
\newcommand\copyrightnotice{%
    \begin{tikzpicture}[remember picture,overlay]
        \node[anchor=south,yshift=10pt] at (current page.south)
        {\fbox{\parbox{\dimexpr\textwidth-\fboxsep-\fboxrule\relax}{\copyrighttext}}};
    \end{tikzpicture}%
}

\begin{document}
\title{\textit{Sparbit}: a new logarithmic-cost and data locality-aware MPI Allgather algorithm}

\author{\IEEEauthorblockN{Wilton Jaciel Loch}
\IEEEauthorblockA{Graduate Program in Applied Computing\\
Santa Catarina State University (UDESC)\\
wilton.loch@edu.udesc.br}
\and
\IEEEauthorblockN{Guilherme Pi\^{e}gas Koslovski}
\IEEEauthorblockA{Graduate Program in Applied Computing\\
Santa Catarina State University (UDESC)\\
guilherme.koslovski@udesc.br}
}

\maketitle

\copyrightnotice

\begin{abstract}

The collective operations are considered critical for improving the performance of exascale-ready
and high-performance computing applications. On this paper we focus on the Message-Passing Interface
(MPI) Allgather many to many collective, which is amongst the most called and time-consuming
operations. Each MPI algorithm for this call suffers from different operational and performance
limitations, that might include only working for restricted cases, requiring linear amounts of
communication steps with the growth in number of processes, memory copies and shifts to assure
correct data organization, and non-local data exchange patterns, most of which negatively contribute
to the total operation time. All these characteristics create an environment where there is no
algorithm which is the best for all cases and this consequently implies that careful choices of
alternatives must be made to execute the call.
Considering such aspects, we propose the Stripe Parallel Binomial Trees (Sparbit) algorithm, which
has optimal latency and bandwidth time costs with no usage restrictions.  It also maintains a much
more local communication pattern that minimizes the delays due  to long range exchanges, allowing
the extraction of more performance from current systems when compared with asymptotically equivalent
alternatives. 
On its best scenario, Sparbit surpassed the traditional MPI algorithms on 46.43\% of test cases with
mean (median) improvements of 34.7\% (26.16\%) and highest reaching 84.16\%.  
\end{abstract}


\IEEEpeerreviewmaketitle

\section{Introduction}
\label{sec:introduction}

The Message Passing Interface (MPI) standard defines several constructs to handle communication
between concurrent processes. These can be generally divided in two forms of message exchanges:
point to point operations, where two processes communicate individually with each other through one-
or two-way data movements; and collective operations, where multiple processes contribute as data
sources, sinks or both into a broader coordinated exchange~\cite{mpi_std_2015}. Although both forms
have significant usage on High-Performance Computing (HPC) applications, the latter is responsible
for the majority of communication time spent on MPI and also accounts for the most number of calls
\cite{chunduri_characterization_2018}. Therefore, the collective operations are considered critical
to improve the performance of parallel and distributed memory systems~\cite{gong_network_2015}. On
this paper, we focus on the Allgather (\texttt{MPI\_Allgather}) many to many operation, which
amongst other collectives holds a large share of utilization~\cite{laguna_large-scale_2019} and time
consumption~\cite{chunduri_characterization_2018} on applications.

The delay magnitude for the completion of a collective is a product of several factors, which
include the underlying hardware topology~\cite{bosilca_online_2017}, communication protocols,
network capacity~\cite{gong_network_2015}, placement of processes~\cite{kurnosov_dynamic_2016,
alvarez-llorente_formal_2017} and others. However, one of paramount importance is the performance
of the algorithm employed to coordinate the high level inter process communication and block
transferences~\cite{benson_comparison_2003, kumar_optimization_2016}. On a purely theoretical
regard, the community has long discovered the minimum time costs to perform this task, both in the
latency term - or the required number of steps - and in the bandwidth term - or the actual data
transfer time~\cite{bruck_efficient_1997}. This, in turn, led to the development of theoretically
optimal algorithms which are capable of performing all the exchanges while maintaining minimum
costs.

For Allgather, the generally available algorithms are Ring, Neighbor Exchange, Bruck and Recursive
Doubling. The Ring algorithm has a linear growth of both latency and bandwidth time with the
increase in the number of processes~\cite{alvarez-llorente_formal_2017}. Neighbor Exchange has the
same asymptotic behaviour but with a less steep increase in time, requiring only half of the Ring's
steps, with the downside of only working for even numbers of
processes~\cite{jing_chen_performance_2005}. Recursive Doubling reaches the minimum costs by
delivering the data with a logarithmic increase in latency, but only works with power of two numbers
of processes~\cite{thakur_optimization_2005}. Finally, Bruck works for any number of processes and
also has a logarithmic cost, requiring only one additional step in non power of two process
numbers~\cite{bruck_efficient_1997}. 

\begin{figure}[!ht]
    \centering
    \includegraphics[width=\columnwidth]{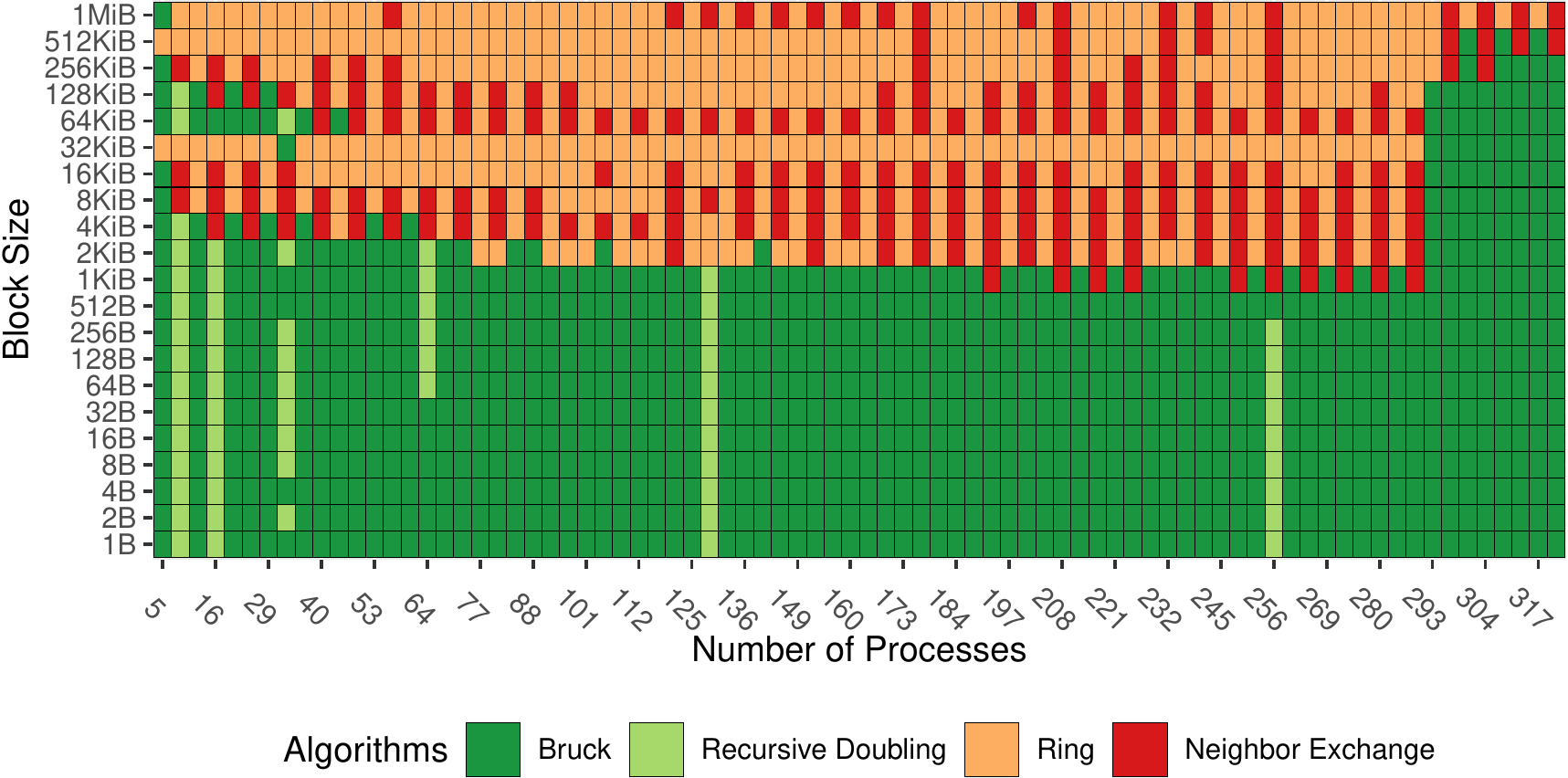}
    \caption{Algorithms with the best experimental time for each case.}
    \label{fig:algorithms_without_sparbit}
\end{figure}

One could expect that for being optimal cost algorithms both Recursive Doubling and Bruck would be
employed at all cases. Nonetheless, the logarithmic options may not yield the best performances when
compared with theoretically less efficient ones as Ring~\cite{thakur_optimization_2005} and Neighbor
Exchange~\cite{jing_chen_performance_2005}, and the reason for these observations lies on practical
limitations that arise when the algorithms are implemented. In practice, Bruck and Recursive
Doubling are only employed for reduced data sizes, because although both have the smallest costs,
their communication patterns are too diffuse, which severely hurts their performance with large data
sizes. As the size of the block to be transferred grows, Ring and Neighbor Exchange are employed
because despite having a worse cost, their communication patterns are local and hence, less
expensive to move data~\cite{thakur_optimization_2005}. Figure \ref{fig:algorithms_without_sparbit}
presents a heat map displaying the algorithm with smallest time for each combination of data size
and number of processes (experiments are further detailed on Section
\ref{sec:experimental_methodology}) and portrays the aforementioned characteristic of performance
degradation for logarithmic algorithms on large data sizes, where linear options achieve the best
times. It is also clear how on power of two numbers of processes Recursive Doubling generally
achieves better times than Bruck, while a similar pattern can be seen for even numbers of processes
with Neighbor Exchange in relation to Ring. All these observations highlight the fact that there is
no \textit{silver bullet} algorithm which is the best for all scenarios and test cases.

On the current collective algorithm scenario, although certain options provide the minimum
theoretical costs, their performance is damaged by non-local communication patterns. This hindering
phenomenon is so severe that it turns linear algorithms into more efficient alternatives, rendering
the advantage of logarithmic number of steps virtually useless in many cases. We therefore propose
the Stripe Parallel Binomial Trees (Sparbit) algorithm, which has logarithmic costs and still
maintains local data exchange patterns. Sparbit was better than the currently employed algorithms
on 46.43\% of test cases (624/1344) with mean (median) improvements of 34.7\% (26.16\%) and highest
reaching 84.16\% on its best scenario.

The additional performance provided by our novel algorithm, which improved upon the state of the art
in several cases, will only be leveraged if an efficient algorithm selection policy is employed.
This however has by itself a consistent dedicated research effort, like the works of
\cite{faraj_star-mpi_2006} and \cite{hunold_algorithm_2018} as well as the Open Tool for Parameter
Optimization (OTPO) focused on Open MPI.
Therefore, we retain our focus towards presenting and validating the proposal, while usage
guidelines are the aim of other independent works.
This paper is organized as follows. Section \ref{sec:algorithms_problem} provides a deeper look into
the currently employed Allgather algorithms. Sparbit and its functioning are introduced on Section
\ref{sec:sparbit}. The experimental methodology is provided on Section
\ref{sec:experimental_methodology} alongside benchmarks, configurations and other important
information. On Section \ref{sec:experimental_analysis} the experimental results are analysed,
Section \ref{sec:related_work} presents the related work and Section \ref{sec:conclusion} concludes
the paper.

\section{Algorithms and problem definition}
\label{sec:algorithms_problem}

The \texttt{MPI\_Allgather} call is a many to many collective operation where each process sends a
fixed size block of data to all other participants and receives their individual blocks as well.
The block sent by a process has one or more items of a chosen datatype and both the size and type
signature of the block must be equal to all destinations. After the Allgather execution, the block
sent by the $j$th process is received by all others and placed in the $j$th position of their
receive buffer~\cite{mpi_std_2015}, whose final configuration will be equal to all processes.
Next, we provide a brief description of the algorithms gathered from the literature and from source
code analysis of both MPICH 3.3.2 and Open MPI 4.0.3. Our focus was directed towards these
implementations since they are open source and the base for other closed source ones. Then, driving
from the presented information and discussion, we formulate the problem.

\subsection{Allgather algorithms} \label{subsec:algorithms}

Initially, a cost model is necessary in order to accurately compare the algorithms that implement
the Allgather operation. To this end we present each algorithm's complexity as cost in time
following the Hockney model \cite{hockney_communication_1994}. With this approach the cost for
sending a message is given by a latency or start-up term $\alpha$ and a bandwidth or transference
cost per byte term $\beta$. 
Since all the presented algorithms essentially implement communication patterns built of individual
point to point messages, the total cost of each algorithm following the Hockney model is given by
the time sum of all the building block messages.%

\paragraph{\textbf{Ring}} \label{subsec:ring}

The Ring algorithm performs the block exchanges by shifting data around on a ring communication
topology. On every step, a process with rank $r$ will send a block to the process with rank $r + 1$
and receive another from the process with rank $r - 1$ (wrapping around if a destination or source
is out of bounds). 
Each process's own block is sent on the first step, while on all others the block received on the
previous step is forwarded~\cite{alvarez-llorente_formal_2017}.
Hereafter, the number of processes involved in the algorithm is represented by $p$, while $m$
represents the total amount of data that a process must have at the end of the operation. In this
sense, each process has to receive a block with size $\frac{m}{p}$ from all the other $p - 1$
participants as well as send its own. Since all sends happen in parallel and any block takes $p - 1$
steps to reach all processes, the cost is given by $C_{ring} = (p - 1)(\alpha + \frac{m}{p}\beta)$.
The ring is employed both on MPICH and Open MPI for various cases.

\paragraph{\textbf{Neighbor Exchange}} \label{subsec:neighbor_exchange}

The Neighbor Exchange is another linear complexity algorithm, however its latency growth is smaller
than the Ring by half. Its main driving idea is to make all transferences pairwise to better
leverage the TCP protocol, such that each process on a step $s$, with $0 \leq s < \frac{p}{2}$, will
only communicate with one peer, both sending and receiving data.
Following this, an even process $r$ will send data to its neighbor $r + (-1)^s$ while an odd process
$r'$ will send data to its neighbor $r' - (-1)^s$, both with wrap-around. 
As two blocks are sent on all steps but the first, this algorithm's cost given by $C_{ne} =
\frac{p\alpha}{2} + (p - 1)\frac{m}{p}\beta$\cite{jing_chen_performance_2005}. The negative side is
that it only works with an even number of processes. The Neighbor Exchange is not employed on MPICH,
but is utilized for several cases on Open MPI.

\paragraph{\textbf{Recursive Doubling}} \label{subsec:recursive_doubling}

The third available algorithm is the Recursive Doubling, which performs the complete exchange in a
logarithmic amount of steps. The main difference of this algorithm in relation to the previous ones
is that it doubles both the distance of the communication and the data being sent.
On every step $s$, with $0 \leq s < log_2p$ a process with rank $r$ will exchange data in a pairwise
fashion with a process that has rank $r \oplus 2^s$, where $\oplus$ represents the binary exclusive
\textit{or}.
As all process send all the data received so far, the number of blocks double at every step and the
cost of the Recursive Doubling is given by $C_{rd} = (log_2p)\alpha + (p -
1)\frac{m}{p}\beta$~\cite{mirsadeghi_topology-aware_2016}.
The operation of this algorithm is limited only to numbers of processes that are powers of two and
thus this is the only case where it is employed both on MPICH and Open MPI.

\paragraph{\textbf{Bruck}} \label{subsec:bruck}

The Bruck is very similar to the Recursive Doubling on its core, doubling distances and data.
However, it does not employ pairwise exchanges and on every step $s$, with $0 \leq s < \lfloor
log_2p \rfloor$, a process $r$ will send all the data it has to the process with rank $r - 2^s$ and
receive new data from the process with rank $r + 2^s$. 
If the number of processes is not a power of two, an additional step is needed where only the first
$p - 2^{\lfloor log_2p \rfloor}$ blocks in the receive buffer are sent~\cite{bruck_efficient_1997}. 
The cost of the Bruck is given by $C_{bruck} = \lceil log_2p \rceil \alpha + (p -
1)\frac{m}{p}\beta$, it has no usage restrictions and is employed both on MPICH and Open MPI for
various scenarios.

\subsection{Problem formulation} \label{subsec:problem_formulation}

So far while presenting the algorithms we have only discussed usage limitations that are inherent to
their architecture. However, in practice the algorithms also have usage limitations that stem from
performance issues, which tend to be side effects of their relation with the supporting
technologies. The formal definitions of the algorithms assume equally balanced communication costs
to all peers, but computing clusters and supercomputers often employ hierarchical network
topologies~\cite{kurnosov_dynamic_2016}. On these networks the cost for performing communication
between two nodes is highly dependent on the physical location of each
peer~\cite{alvarez-llorente_formal_2017}, and the further away they are, the longer are the physical
paths between them and therefore the higher the latency. From a bandwidth perspective, the further
away two nodes are the higher is the chance that their communication will cross the core of the
network, whose bandwidth is more expensive and supports less saturation than the
edge~\cite{wang_locality-aware_2018}, possibly leading to slowdowns or contentions. 

Along these lines, although Recursive Doubling and Bruck have a smaller number of steps, they
naturally communicate over longer paths on the network, implying in higher time costs specially for
larger messages. Additionally, Bruck also needs memory shifts to assure contiguous sends and data
organization, which further deteriorate its performance with the increase in data size. On the
other hand, Ring and Neighbor Exchange have worse linear latency costs, but have local communication
patterns always with the same neighbors, which implies in reduced additional transfer costs with the
growth of message size~\cite{thakur_optimization_2005}.

This scenario outlines a conflict between the theoretical quality of the algorithms and their actual
efficiency: so far, the performance of logarithmic algorithms is degraded by their non-local
communication to such an extent that they achieve worse times than linear options in many cases.
This phenomenon hinders their potential for reduced time costs and limits the overall application
performance.

\section{The Sparbit algorithm} \label{sec:sparbit}

This section outlines the design of the Stripe Parallel Binomial Trees (Sparbit) algorithm.
It is inherent to the logarithmic communication algorithms that the amount of data exchanged at any
given step will be larger than the amount of the previous one, usually by a factor of two, implying
that the cost of data exchange grows at every step as a product of data size. This is true for
Bruck, Recursive Doubling and syllogistically for Sparbit too. 
Additionally, if we suppose the usage of hierarchical network topologies, then the cost of sending
and receiving data increases with the growth of distance between any two communicating peers.
We can hence derive a cost estimate for every step as being a product of these two dimensions: the
larger the data size and the further the distances travelled by the data, the higher will be the
cost in time for the communication. The former dimension can not be changed in order to maintain the
logarithmic complexity, however there is no restriction on the latter. Stemming from this and
through a higher behaviour perspective, the goal of Sparbit is to provide the same data exchange
semantics as the Bruck algorithm, however instead of doubling the distances starting from $1$ until
$2^{\lceil log_2p \rceil - 1}$, it takes the opposite direction and halves them starting from
$2^{\lceil log_2p \rceil - 1}$ until $1$. This creates a more balanced distribution of communication
costs along the execution, since as the data sizes inevitably grow, the distance they must traverse
continuously shrinks. 

\subsection{Binomial tree} \label{subsec:binomial_tree}

\begin{figure}[ht] \centering
\includegraphics[width=.8\columnwidth]{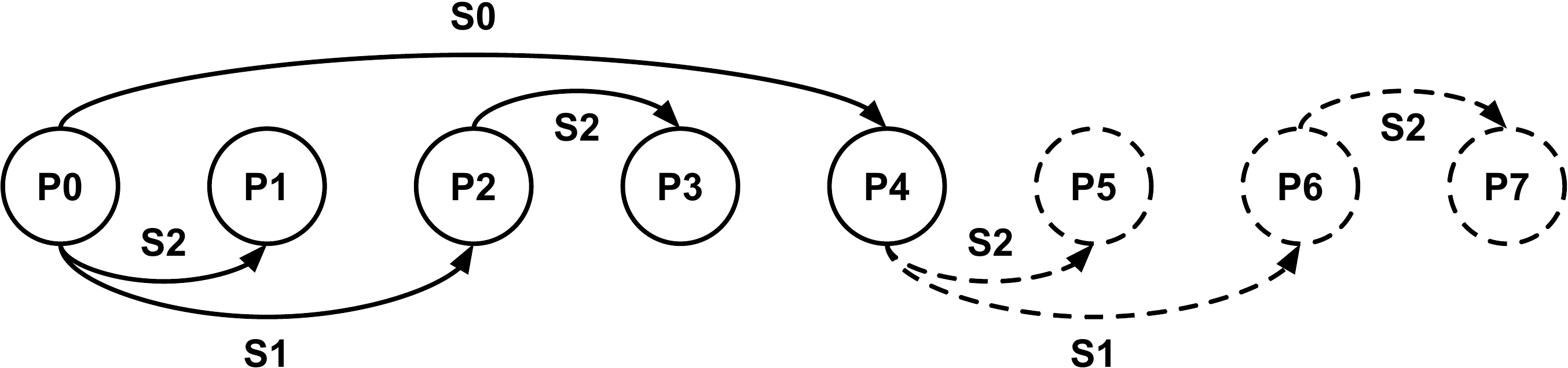} \caption{Binomial
tree. Dashed circles and arrows represent missing processes and ignored sends, respectively.}
\label{fig:binomial_tree} \end{figure}

As suggested on its name, Sparbit uses the binomial tree to deliver data, which is a one to many
logarithmic data delivery algorithm. Initially on its execution (Figure \ref{fig:binomial_tree}),
the block of data resides in the root process, here assumed to have rank~$0$. On the first step,
the root will send the block to the process with rank $p/2$ (P4) and will delegate to it the
responsibility of delivering the data to the processes with rank $r > p/2$. Therefore, this first
step has divided the group of processes in two subtrees, with each root $0$ and $p/2$ responsible
for sending the block to half of the processes. The algorithm is then repeated recursively and in
parallel for each of the halves, continuously subdividing the processes into smaller subtrees until
the data has reached all of them. 

For a non power of two number or processes, as exemplified with 5 on Figure \ref{fig:binomial_tree},
the algorithm works by exactly mimicking the way a binomial tree with size $2^{\lceil log_2p
\rceil}$ would work, however ignoring any sends that are destined to processes with ranks equal or
greater than $p$. Since all sends happen in parallel, the cost of the binomial tree algorithm is
given by $C_{binomial\_tree} = \lceil log_2p \rceil (\alpha + m\beta)$.

\subsection{Algorithm design} \label{subsec:algorithm_design}

Regarding the semantics of data delivery on an MPI Allgather call with $p$ participants, each
process must somehow send its block to all the other $p - 1$ peers. If we take only the process with
rank $0$ and suppose that it is the root of a binomial tree composed of all the participating
processes, then it would be trivial to deliver its block to all the destinations in a logarithmic
number of steps.
Moreover, if we imagine that the process rank space is circularly connected (\textit{i.e.}, there is
wrap-around on the destination calculation), then we can expand our previous supposition to any
process $r$, since the tree organization would be exactly the same only shifted by $r$ units.
Finally from this concept, if all processes act as roots of the binomial tree destined to distribute
their own block and simultaneously as leafs or intermediate nodes of all the other $p - 1$ trees,
the whole Allgather semantic can be achieved. This idea is the core of Sparbit and virtually all
that it does is start and manage the participation of each process in the execution of all these
parallel binomial trees. 
The resulting communication pattern of Sparbit can be also modelled as a Binomial
Graph~\cite{angskun_binomial_2007}, however utilizing a unique data distribution behaviour.
Bruck employs a similar general concept but with inverse distance circulant graphs, which manifest
themselves in the form of parallel spanning trees that cause its locality issues.

Although the idea of Sparbit is simple, there is a considerable disconnect between its observed
behaviour and the actual algorithmic logic that makes it work. On the execution of the binomial tree
distribution, on each step with distance $d$ a process that already has the block will send it to a
peer $d$ away on the process rank space. 
If the number of processes is a power of two, then the execution is trivial and on every step s,
with $0 \leq s < log_2p$, a process both sends and receives $2^s$ blocks to and from destinations
$2^{log_2p - (s + 1)}$ away on each direction. 
However, if the amount of processes is not a power of two, then some block sends will be ignored to
avoid unexisting destinations. When employing wrap-around there are no unexisting destinations, but
facing the same execution scenario and not ignoring these sends would imply in double writing the
same block, which is not desirable. Hence, if we take again the example shown in Figure
\ref{fig:binomial_tree} with 5 processes and imagine that all the blocks initially reside on process
0, then on the first step process 4 would receive them. Now, since process 4 is a leaf and thus has
no destinations, it would simply ignore the regular behaviour of forwarding the received blocks.
When looking to the real Sparbit scenario the same logic applies, however the blocks are scattered
along all the processes, with each distributing its portion through its own individual tree. Since
each tree is shifted towards a unique root and all processes participate in all trees, then each
process must act as the relative fourth destination - $(root + 4) \% p$ - to one of these trees.
This implies that the role of ignoring sends, done in the initial supposition only by process 4, now
must be done by all processes, each ignoring the blocks received through the trees in which it is a
leaf, but still forwarding the ones from trees in which it is an intermediate node.

\begin{figure}[ht] \centering
\includegraphics[width=\columnwidth]{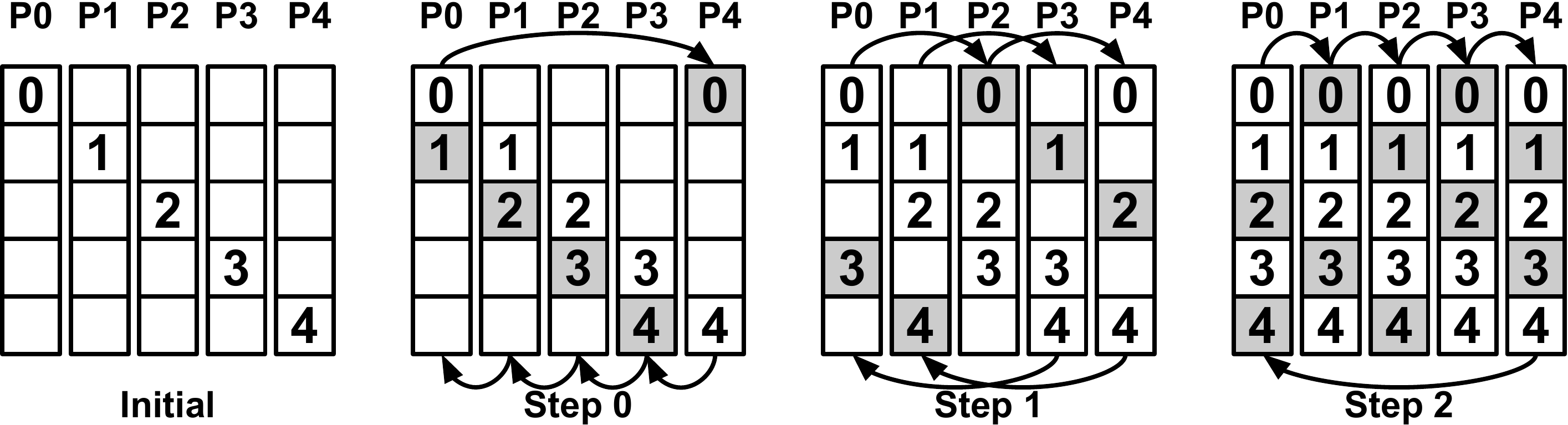} \caption{Sparbit
buffer configuration with 5 processes on different steps. Grey blocks are received on the current
step.} \label{fig:sparbit_buffer_configuration} \end{figure}

The collective ignoring and forwarding behaviour can be clearly observed on Figure
\ref{fig:sparbit_buffer_configuration}, that shows the buffer configuration for each process on
different steps of Sparbit's execution. On step 0, each process $r$ both sends its original block to
process $r + 4$ and receives the original block of process $r - 4$, with wrap-around. Since the
received block comes from the left, it is placed on the corresponding upwards distance on the
receive buffer, also with wrap-around as can be seen on processes 0 to 3. On the beginning of step 1
all processes have two blocks, however since the one received on the previous step comes from trees
where the processes are leafs, it must not be forwarded and again only the original block is sent on
this step with distance 2. On the last step, each process has 3 blocks and the same one must still
be ignored. Nonetheless, the other two belong respectively to the tree in which the process is a
root and to the one where it is an intermediate node, thus both must be forwarded with distance one.

\begin{algorithm}
\SetAlgoLined\DontPrintSemicolon
\SetKwInOut{Input}{input}
\Input{The amount of processes $p$; the rank of the current process $rank$}
    $data \gets 1$;\\
    $ignore \gets 0$;\\
    $d \gets 2^{\lceil log_2p \rceil - 1}$;\\
    $last\_ignore \gets count\_trailing\_zeros(p)$;\\
    $ignore\_steps \gets (\neg(p\ >>\ last\_ignore)\ |\ 1)\ <<\ last\_ignore$;\\
    \For{$0 \leq i < \lceil log_2p \rceil$}{
        \If{$d\ \&\ ignore\_vector$}{
            $ignore \gets 1$;\\
        }
        \For{$0 \leq j < data - ignore$}{
            \texttt{MPI\_Isend}($rank + d\ \%\ p$, $(rank - (2j)d + p)\ \%\ p$, 1);\\
            \texttt{MPI\_Irecv}($(rank - d + p)\ \%\ p$, $(rank - (2j + 1)d + p)\ \%\ p$, 1);\\
        }
        \texttt{MPI\_Waitall()}\\
        $d \gets d >> 1$;\\
        $data \gets (data << 1) -  ignore$;\\
        $ignore \gets 0$;\\
    }
\caption{Sparbit.}
\label{alg:sparbit}
\end{algorithm}

Another way to see a binomial tree is as a recursive structure composed of smaller power of two
binomial trees, whose roots must receive the data to further distribute it amongst their branches.
The amount and size of subtrees is respectively defined by the number and values of the powers of
two which compose $p$. In the example of Figure \ref{fig:binomial_tree} with 5 processes there are
two such subtrees, one with 4 processes of ranks 0 to 3, and another with 1 process of rank 4.
Therefore if we employ this rationale on the example, the first step takes the data from the root of
the first tree to the root of the second one, performing what we call a tree expansion. If the
number of processes was larger there would be more subtrees, and therefore more tree expansions
during the execution. For example, if there were 21 processes, there would be three subtrees of
sizes 16, 4 and 1. Also since the distance halves, the trees are organized in descending order, with
larger subtrees comprising the smaller relative ranks and the global root always being the root of
the first one. Therefore, the block flows from the larger to the smaller subtrees and a tree
expansion will only happen on a step with distance $d$ if there is a subtree of size $s_1 = d$ that
has the block and another subtree with size $s_2 < s_1$ that does not. 
Practically, a tree expansion only indicates that on that step every process should send all the
blocks it has.

\begin{figure} \centering \includegraphics[width=.9\columnwidth]{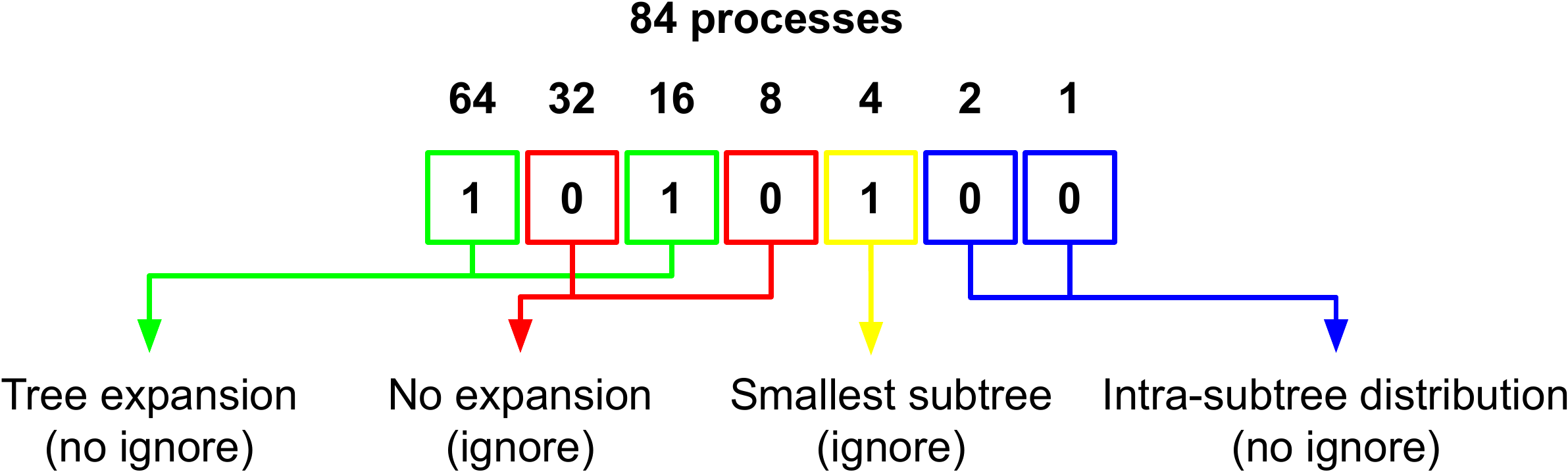}
\caption{Binary representation with behaviour on each bit.} \label{fig:binary_representation}
\end{figure}

With all this in mind, Algorithm~\ref{alg:sparbit} presents Sparbit's pseudocode. 
Function signatures for \texttt{Isend} and \texttt{Irecv} were simplified for brevity, taking
destination/source, buffer displacement in blocks and number of blocks as parameters.
The expected amount of data and whether the sending of a block should be ignored on the current step
are kept in the \texttt{data} and \texttt{ignore} variables, respectively. 
The \texttt{ignore\_steps} variable stores the steps in which a block should be ignored, being built
from the binary representation of $p$ as shown on \Cref{fig:binary_representation} for an arbitrary
number.
For $t$ subtrees, since the data is already initially located on the first one, only $t-1$
expansions are needed. These expansions happen on the $t-1$ larger subtrees and thus no expansion
happens on the smallest one. Bellow the smallest subtree all blocks should be sent on every step as
now all distributions are local. 
Such patterns are highlighted on \Cref{fig:binary_representation} and if we invert all bits to the
left of the first one set, then steps in which a block must be ignored will have their respective
bits set, while unset bits will indicate steps with the opposite behaviour. 
All these binary operations are performed on lines 4 and 5 of \Cref{alg:sparbit}. Through $log_2p$
steps (lines 6-18) the proper blocks get sends and receives issued (lines 10-13) and waited for
(line 14), the distance halves as the data size doubles (lines 15 and 16). If in the current step a
block must be ignored, the proper variable is set (lines 7-9) and the number of blocks sent is
reduced (line 10). Since all sends happen in parallel and Sparbit takes a logarithmic number of
steps, its cost is given by $C_{sparbit} = \lceil log_2p \rceil\alpha + (p - 1)\frac{m}{p}\beta$.

\subsection{Process mappings, topology and Sparbit} \label{subsec:process_mappings}

The mapping of MPI processes to the topology is done by assigning ranks to their instantiations on
the machines. Naturally, since the communications are established by means of process' ranks, the
manner in which they are distributed plays a significant role on the respective costs. 
The two main predefined ways in which MPI implementations map ranks to nodes are \textit{sequential}
and \textit{cyclic}. The first performs a best-fit approach, completely filling the slots of a
machine with sequential ranks before going to the next. The second uses a round-robin approach,
assigning only one rank to a machine and then moving to the next in a circular fashion until they
are all filled. By default, Open MPI employs sequential mapping while MPICH employs
cyclic~\cite{alvarez-llorente_formal_2017}.

Sparbit is expected to achieve greater locality under sequential mapping, as the heavier
communication stages are held among closer ranks, which in this case are physically mapped near to
each other. Meanwhile, Bruck is expected to perform better under cyclic mapping due to the same
reason.

\section{Experimental methodology} \label{sec:experimental_methodology}

Two sets of experiments were executed on the Yahoo and Cervino clusters from the University of
Neuch\^{a}tel, where 16 and 5 dedicated homogeneous machines were employed, respectively. On the
Yahoo tests, the network is organized in a two-tier tree topology with 5 machines connected to one
Gigabit Ethernet leaf switch, 11 machines connected to another equal one, and both interconnected by
a core switch through individual 10Gbps single links. Every machine has 8 processing cores provided
by two Intel Xeon L5420 CPUs, each containing 4 cores and 1 thread per core. The available memory
on each node is 8GiBs. On the Cervino tests, the network is organized in a flat topology with all
nodes connected directly to the same switch via 40Gbps individual links. Every machine has 32
processing cores provided by two Intel Xeon E5-2683 v4 CPUs, each containing 16 cores and 2 threads
per core. The available memory on each node is 128GiBs. For both experiments the employed operating
system was Ubuntu 20.04.2; the chosen MPI implementation was Open MPI 4.0.3; and the benchmark
employed was the OSU Micro benchmarks on version 5.7, executing the collective test
\texttt{osu\_allgather}. Although OSU is an opaque benchmark~\cite{stanisic_characterizing_2017}, we
have chosen it due to its wide utilization on MPI communication research.

\subsection{Metrics and configurations} \label{subsec:metrics_configurations}

All metrics were collected through the OSU reports, namely minimum, maximum and average time of the
Allgather call over all runs. The Sparbit algorithm for \texttt{MPI\_Allgather} was developed in C
and bundled into an Open MPI collective communication component. The compared algorithms were
Bruck, Recursive Doubling, Ring and Neighbor Exchange. Each algorithm was selected for execution
through command line arguments passed to the \texttt{mpirun} executable, with native algorithms
chosen through the dynamic rules of the \texttt{tuned} component, and Sparbit selected by raising
its component priority.
Block sizes were varied following OSU's default limits and variation, starting at 1B and raising up
to 1MiB on successive multiplications by 2, totalling 24 tested data sizes.

The maximum number of processes utilized on the Yahoo tests was 256, while on Cervino it was 320,
respecting a limit of 2 processes per physical core of each machine for both infrastructures. The
processes were varied following two arithmetic progressions, both with step difference of 8, the
first starting at 8 and ending at 256 for Yahoo (320 for Cervino) to represent cases of even and
power of two numbers of processes. The second starting at 5 and ending at 253 (317) represented
cases of odd numbers of processes, where there is asymmetry on the algorithms' execution and on the
mapping of processes to the machines. On total there were 64 (80) different numbers of processes,
resulting in 1344 (1680) test cases from combinations with data sizes.
Each singular test of the previous parameters was executed employing both sequential and cyclic
mappings, allowing the empirical validation of the discussion on
Section~\ref{subsec:process_mappings}. In order to reduce the probability and effect of any
unexpected machine or network events on the time of a test, all algorithms were sequentially
executed with both mappings for each number of processes. Finally, each test was executed 50 times
for statistical significance, with 5 additional warm-up executions to reduce the effects of any
possible start-up skews. 

\section{Experimental analysis}
\label{sec:experimental_analysis}

\begin{figure*}[!ht]
    \centering
        \begin{subfigure}[h]{.485\textwidth}
            \centering
            \includegraphics[width = \textwidth]{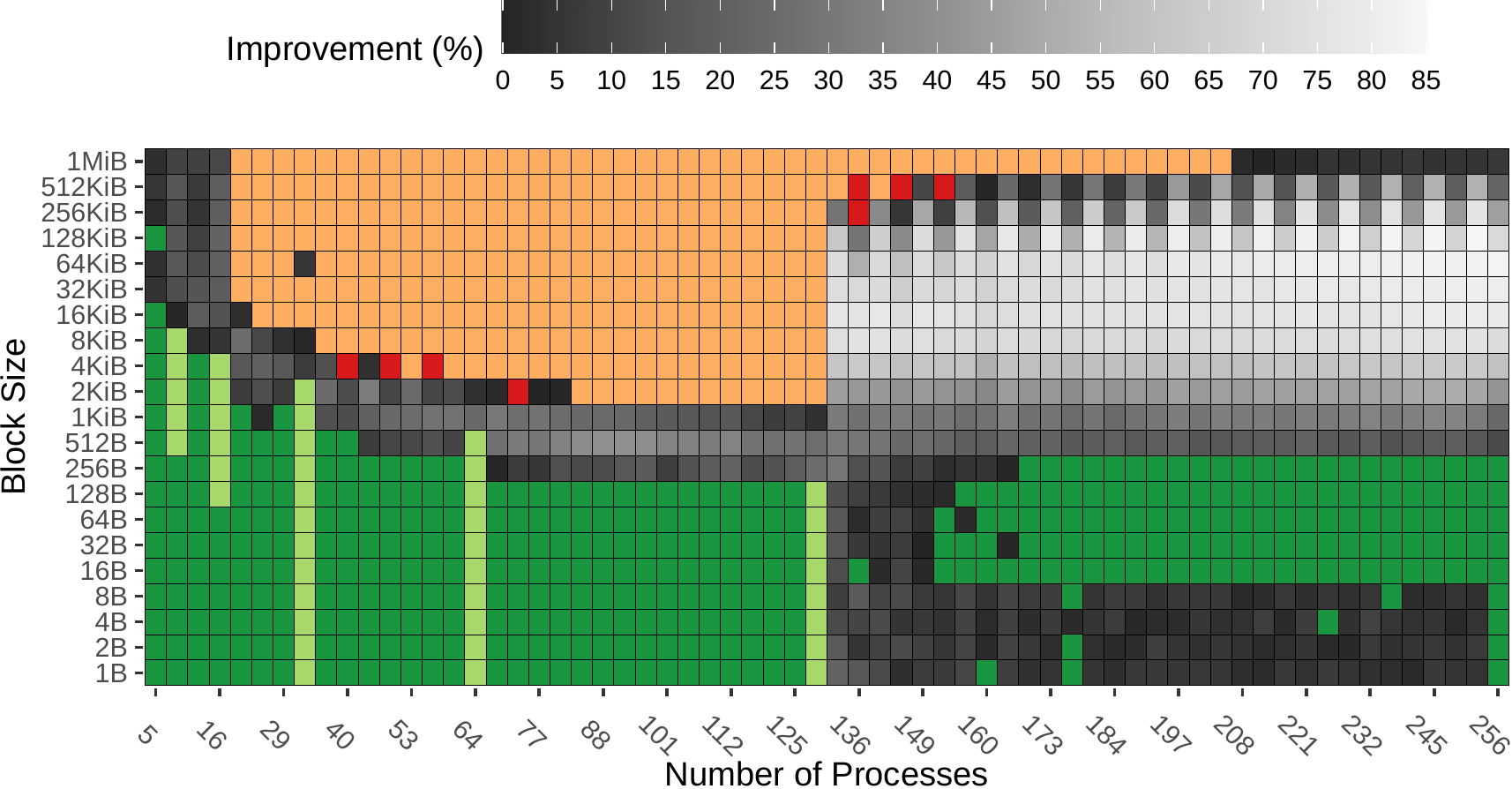}
            \caption{}
            \label{subfig:yahoo_sequential}
        \end{subfigure}
        \begin{subfigure}[h]{.477\textwidth}
            \centering
            \includegraphics[width = \textwidth]{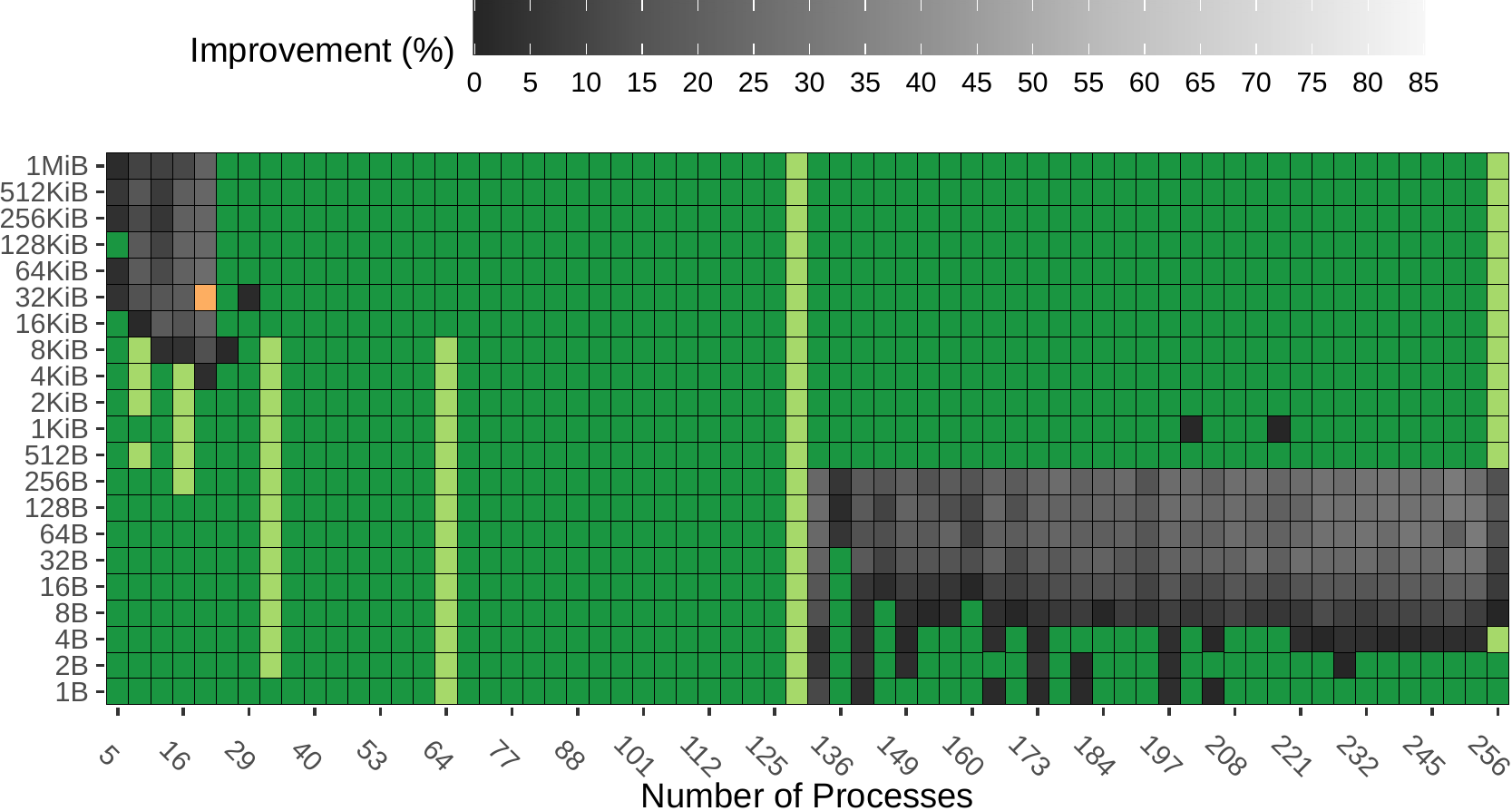}
            \caption{}
            \label{subfig:yahoo_cyclic}
        \end{subfigure}

        \begin{subfigure}[h]{.485\textwidth}
            \centering
            \includegraphics[width = \textwidth]{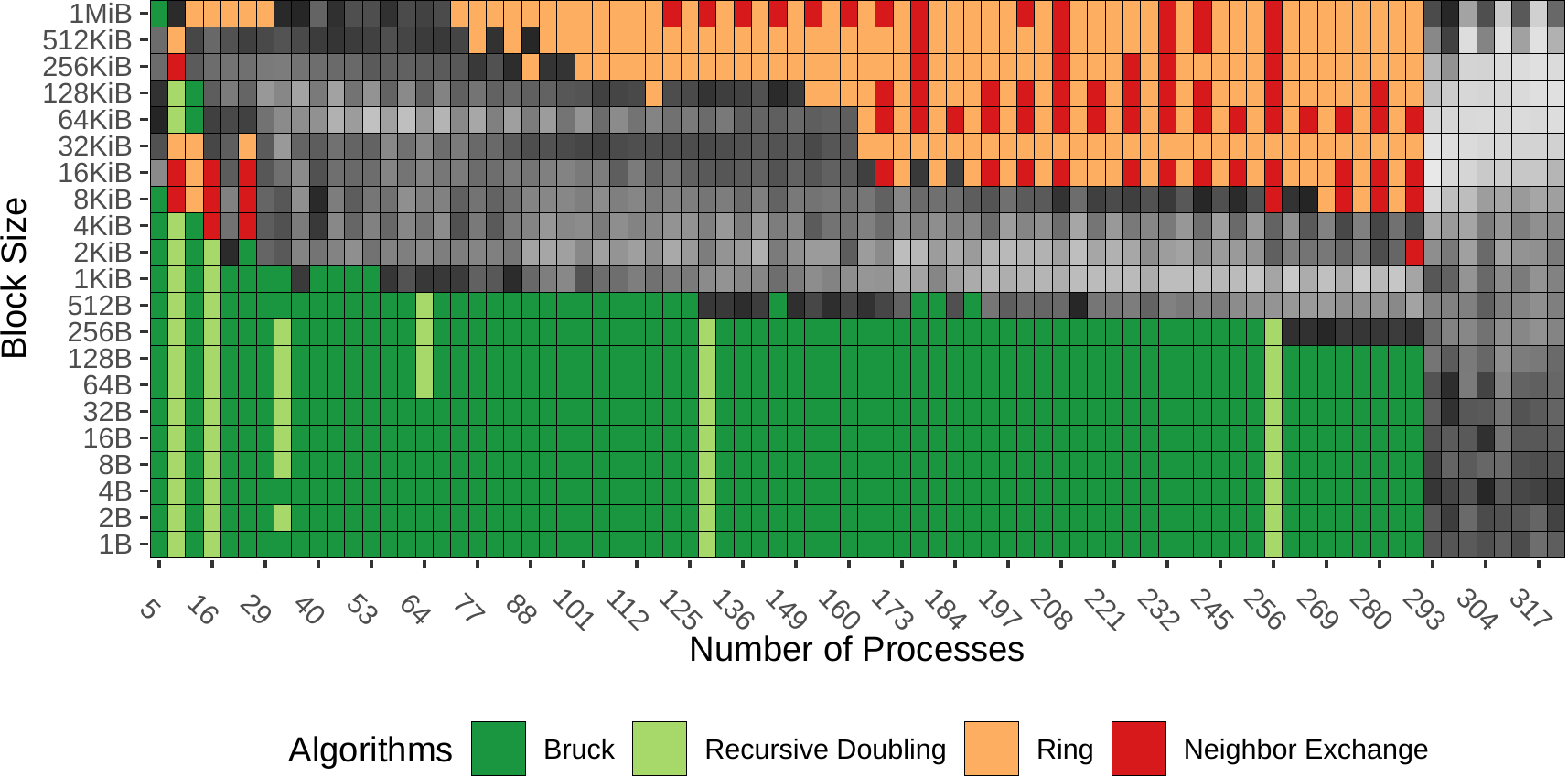}
            \caption{}
            \label{subfig:cervino_sequential}
        \end{subfigure}
        \begin{subfigure}[h]{.477\textwidth}
            \centering
            \includegraphics[width = \textwidth]{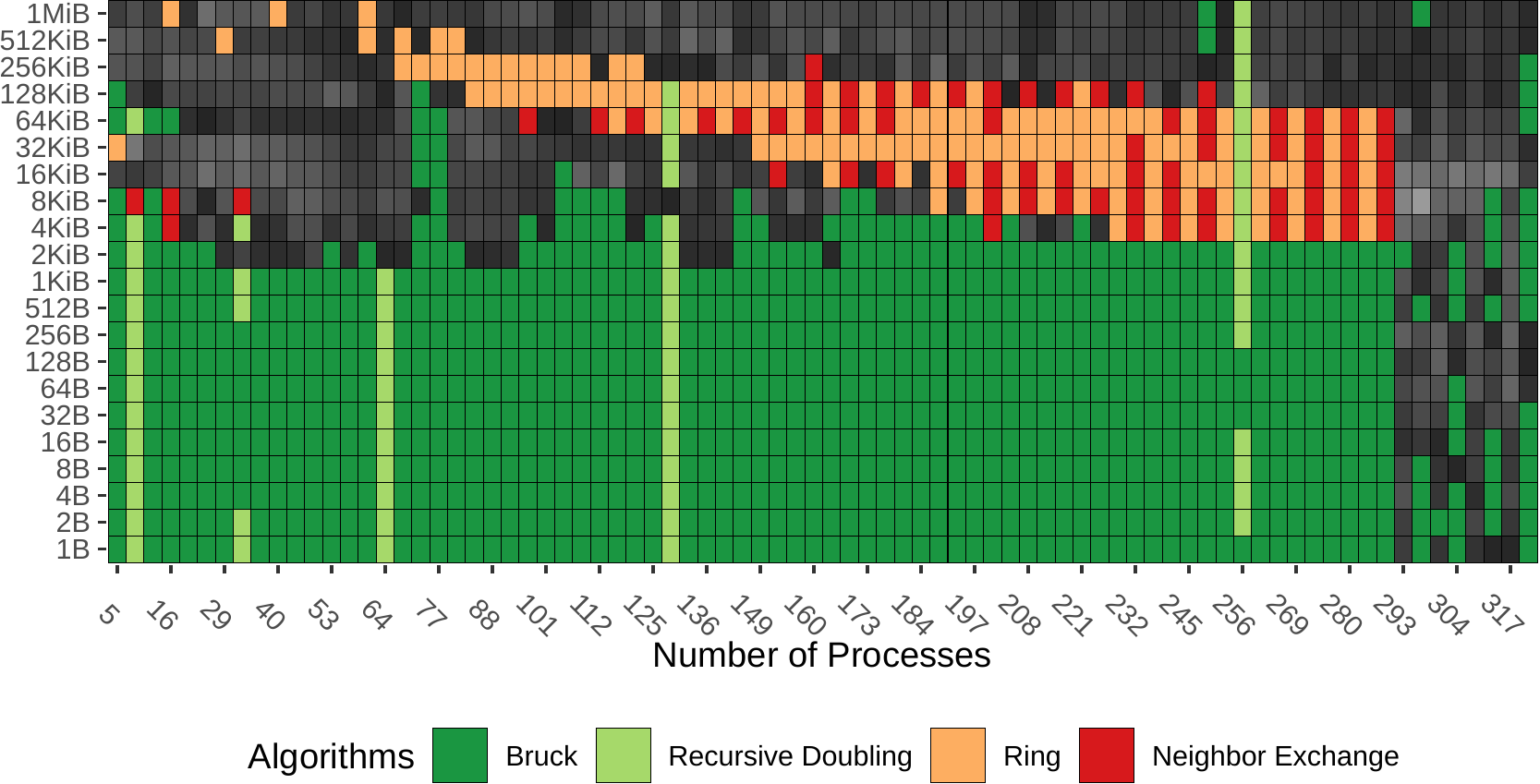}
            \caption{}
            \label{subfig:cervino_cyclic}
        \end{subfigure}
        \caption{Heat maps of best average time algorithm for each case. (a) Yahoo cluster with
        sequential mapping. (b) Yahoo cluster with cyclic mapping. (c) Cervino cluster with
    sequential mapping. (d) Cervino cluster with cyclic mapping.}
    \label{fig:average_heatmaps}
\end{figure*}

\begin{table*}[h!]
\centering
 \begin{tabular}{llrrrrrrrr} 
     Cluster & Mapping (Total) & Min & Avg & Max & Min $\cap$ Avg & Min $\cap$ Max & Avg $\cap$ Max & Min $\cap$ Avg $\cap$ Max \\ 
 \hline
     \multirow{2}{*}{Yahoo} & Sequential (746) & 115 (15.42\%) & 5 (0.67\%) & 6 (0.8\%) & 113 (15.15\%) & 1 (0.13\%) & 14 (1.88\%) & 492 (65.95\%) \\ 
                            & Cyclic (346)     & 48 (13.87\%) & 0 (0\%) & 35 (10.12\%) & 5 (1.45\%) & 6 (1.73\%) & 33 (9.54\%) & 219 (63.29\%) \\
 \hline
     \multirow{2}{*}{Cervino} & Sequential (799) & 133 (16.65\%) & 6 (0.75\%) & 0 (0\%) & 196 (24.53\%) & 0 (0\%) & 3 (0.38\%) & 461 (57.7\%) \\
                              & Cyclic (738) & 202 (27.37\%) & 20 (2.71\%) & 16 (2.17\%) & 92 (12.47\%) & 2 (0.27\%) & 115 (15.58\%) & 291 (39.43\%) \\
 \end{tabular}
 \caption{Relation of Sparbit's best minimum, average and maximum time sets.}
 \label{tab:improvement_sets}
\end{table*}

On this section we present the results obtained from the experiments and the analysis of their data.
Initially, four hybrid heat maps with a broad view of the experiments are provided on Figure
\ref{fig:average_heatmaps}, two for Yahoo and two for Cervino, showing sequential and cyclic cases.
Every cell of each heat map indicates a particular test configuration, comprising a unique
combination of block size and number of processes. Each row indicates one block size from 1B to 1MiB
and each column indicates one process amount, with both arithmetic progressions intertwined and
starting from the odd one (as discussed on Section~\ref{subsec:metrics_configurations}). The heat
maps are said to be hybrid because they present both categorical and continuous values: if Sparbit
does not have the smallest time on a test, the color of the cell indicates what traditional
algorithm was the best (\textit{e.g.}, the dark green color for 1B and 5 processes on Figure
\ref{subfig:yahoo_cyclic} indicates that Bruck was the best on this configuration). In turn, if
Sparbit has the smallest time, a lighter is better grey scale color indicates its improvement
percentage over the second best algorithm -- \textit{e.g.}, 23.83\% improvement for 1MiB and 320
processes on Figure \ref{subfig:cervino_sequential}. All the heat map values are related to the
smallest average time over 50 executions.

\paragraph{\textbf{Network topology and data locality impact}}

Most of the results that can be seen on Figure \ref{fig:average_heatmaps} corroborate with the
previous discussions on Hockney-based algorithm costs, presented on Section~\ref{subsec:algorithms}.
Under sequential mapping on both infrastructures, Sparbit has the smallest execution time on a large
portion of the tests, reaching 46.43\% on Yahoo and 39.64\% on Cervino, as seen on
Figures~\ref{subfig:yahoo_sequential} and~\ref{subfig:cervino_sequential}, respectively. Under
cyclic mapping there is a considerable drop on Sparbit's percentage, as expected from its worse
communication pattern on this case, reaching 19.12\% on Yahoo and 30.83\% on Cervino, as seen on
Figures~\ref{subfig:yahoo_cyclic} and~\ref{subfig:cervino_cyclic}, respectively. 
The different topologies of the infrastructures give valuable insights on its impact over Sparbit
and the results itself. First, Yahoo has a larger topology with more levels crossing switches, which
implies on different communication costs among distinct pairs of machines.
Cervino on the other hand has a flat topology, which implies on equal theoretical communication
costs among any two machines, leaving only the internal machine communication hierarchy and its
difference to the external one. 
Therefore, on Yahoo with sequential mapping we can expect the highest benefit from employing
Sparbit, which can be seen true by the largest fraction of the best times achieved on this
experiment. The smaller topology of Cervino reduces the maximum potential benefit of Sparbit, which
correspondingly reduces its fraction of the best times. Under cyclic mapping, Sparbit's drop in
performance is expected to be proportional to the topology's size, since now there is an inverted
communication pattern. This observation justifies the drop in relation to the sequential mapping
seen on the Yahoo cluster to be larger than the one seen on Cervino, as the negative impact of the
former's topology is higher than the latter. From all this we can state that the larger the
topology's hierarchy, the larger the potential benefit of using Sparbit under sequential mapping,
and the larger its benefit's reduction when switching to cyclic.

\paragraph{\textbf{Segmented analysis of algorithms' relation}}

The form of the area taken by the Sparbit's best cases (Figure~\ref{fig:average_heatmaps}), or the
particular intervals where it is the best, assume these uncommon shapes due to the cropped nature of
the heat map. If we individually compare the area of smaller groups of algorithms with Sparbit, like
the ones with linear or logarithmic complexity, then more comprehensible forms appear. 
Through this rationale, the forms of Sparbit on the Yahoo and Cervino sequential heat maps are the
product of the same phenomenon. Considering Yahoo with sequential
mapping~(\Cref{subfig:yahoo_sequential}) and stripping Bruck and Recursive Doubling out of the plot,
their area would be almost completely taken by Sparbit, absorbing 95.48\% of the cases as the best,
with Ring and Neighbor Exchange absorbing only the remaining percentage.
If we instead stripped out Ring and Neighbor exchange while maintaining the rest, a similar pattern
would emerge with Sparbit taking 100\% of their area as the best.
These observations show how there is a clear improvement pattern of Sparbit towards each of the
algorithm groups. In relation to Ring and Neighbor Exchange, Sparbit is better on large numbers of
processes and large data sizes (hereafter considered the ones larger than 1KiB), while it is better
on almost all tests (97.44\%) for smaller sizes (hereafter considered the ones smaller or equal than
1KiB). In relation to Bruck and Recursive Doubling, Sparbit is better for large data sizes also on
almost all tests (97.16\%). Therefore on the merged final version, which is shown on Figure
\ref{subfig:yahoo_sequential}, for 128 processes or less Sparbit excels on the verge sweet spot of
both linear and logarithmic class algorithms, where the data size is to small to compensate the
additional steps of the former and to large to compensate the non-local communication of the latter.
Above 128 processes, which can be seen as a behavioural turning point, Sparbit takes over the
majority of the large data sizes as the best - since it is the case on both the stripped versions -
and a fraction of the small ones as well. 

The same general pattern of Yahoo under sequential mapping can also be seen on Cervino. The
difference is that the verge sweet spot for each number of processes is larger (\textit{i.e.},
involves more test cases) and the behavioural turning point appears on 288 processes. We assume that
theses abrupt changes of behaviour are due to the machines entering an overbooked scenario where
there are more process preemptions. On Yahoo, the first load where there are more processes than
cores is with 133, and Cervino by having two threads can suppress this problem until higher ratios,
but also fails before reaching 2 processes per core. This negatively affects all algorithms but is
less severe on Sparbit, particularly on large data sizes, suggesting that it has less trouble when
operating on highly utilized or restricted CPU environments.

Under cyclic mapping, Ring and Neighbor Exchange also suffer a drop in performance, since their
neighbor ranks on the sequential mapping are now placed at other possibly far machines, which
results in a very non-local communication pattern. Regarding Sparbit, on Cervino with cyclic
mapping the area where it is the best results from a similar merging phenomenon as discussed on the
sequential mapping. Here, the difference is that it is better on more cases of large data sizes
previously taken by Ring and Neighbor Exchange, which are surpassed due to their more expressive
drop on performance. On the other hand, it also loses more cases of both small and large data sizes
to Bruck, due to the latter's improved communication locality on this mapping. On Yahoo with cyclic
mapping, the form taken by Sparbit as the best virtually becomes a product of its relation with
Bruck and Recursive Doubling, as it would on some of the previously discussed stripped versions of
the plot. Sparbit is the best on the small top left area due to its use of parallel sends, employing
more available bandwidth sooner, thus resulting in smaller communication time.

\paragraph{\textbf{Smallest minimum and maximum times}}

The additional metrics for smallest minimum and maximum times are discussed in relation to the
previous considerations. The forms taken by Sparbit as the best on the minimum and maximum heat maps
are very similar to the ones taken on the already shown plots of average. Therefore, instead of
presenting new heat maps we recur to Table \ref{tab:improvement_sets}, which presents the relations
amongst the minimum, average and maximum sets comprising cases where Sparbit has the best time.

For each infrastructure, Table \ref{tab:improvement_sets} presents each mapping with the union of
all cases in which Sparbit was the best. The amount of cases with exclusively minimum, average or
maximum smallest time are also presented on the third, forth and fifth columns, respectively. From
columns 6 to 8 the binary set intersections are shown and column 9 indicates the cases on which
Sparbit is simultaneously the best on the three sets. Initially, focusing only on the sequential
mapping, Sparbit has either the minimum, average or maximum best time on 746 (55.51\%) cases for
Yahoo and 799 (47.56\%) for Cervino. An important behaviour to notice is that for both
infrastructures on this mapping there is little to no cases of exclusively average or maximum best
times. This indicates that these metrics are correlated and that, as can be seen on the last two
columns, on the majority of cases where it has the smallest maximum time, it also has the smallest
average time -- 98.64\% (506/513) on Yahoo and 100\% on Cervino. On both infrastructures the
largest set of Sparbit's best times is the minimum, which also accounts for the largest exclusive
number of cases and individual intersections with the average. This indicates that under sequential
mapping it is easier for Sparbit to lower the minimum and average times than it is to lower the
maximum ones. Such observation is further based on the decreasing amount of best cases from minimum
to average and then maximum sets. The intersection of the three sets accounts for 36.61\%
(492/1344) of the total tests on Yahoo and 27.44\% (461/1680) on Cervino, indicating that on these
fractions it performs better than the other algorithms in a consistent way.

Under cyclic mapping, the total number of cases where Sparbit is the best in any of the metrics
drops to 346 (25.74\%) for Yahoo and 738 (43.93\%) for Cervino, which follows the decay pattern seen
previously in detail for the average. Here, the cases are more evenly spread along exclusive and
binary intersections, especially on Cervino. The correlation between average and maximum metrics
still exists but weaker, with 86\% (252/293) and 95.75\% (406/424) of the maximum best cases also
being the best average ones for Yahoo and Cervino, respectively. On the latter, the minimum
exclusive cases are very expressive, probably due to the fastest and more capable machines which
amplify Sparbit's ability to reduce the minimum metric. There is no clear ordering of set sizes that
applies to both infrastructures under this mapping and the intersection of all metrics drops to
16.29\% (219/1344) of the total cases for Yahoo and 17.32\% (291/1680) for Cervino.

\paragraph{\textbf{Sparbit's time reduction}}

So far, the analysis has only focused on the number of cases in which Sparbit performed better than
other algorithms, without discussing the actual improvement obtained on the tests. Table
\ref{tab:improvement_values} summarizes the mean, median and highest percentage improvement values
obtained on all three measured metrics for the different experimental scenarios. We also refer back
to the heat maps of Figure \ref{fig:average_heatmaps} in order to discuss the patterns of
improvement for the proposal.
\begin{table}[h!]
\centering
 \begin{tabular}{llllrrr} 
     Mapping & Cluster & Metric & Mean & Median & Highest \\
     \hline
     \multirow{6}{*}{Sequential} & \multirow{3}{*}{Yahoo} & Min & 36.18 & 29.98 & 85.87 \\ 
                                 &                        & Avg & 34.70 & 26.16 & 84.16 \\ 
                                 &                        & Max & 35.48 & 27.24 & 80.85 \\ 
                               \cline{2-6}
                               & \multirow{3}{*}{Cervino} & Min & 33.08 & 31.69 & 80.20 \\ 
                               &                          & Avg & 30.23 & 29.00 & 77.78 \\ 
                               &                          & Max & 27.67 & 26.14 & 74.27 \\ 
     \hline
     \multirow{6}{*}{Cyclic} & \multirow{3}{*}{Yahoo}     & Min & 13.35 & 12.38 & 42.31 \\ 
                                     &                    & Avg & 14.89 & 15.77 & 31.07 \\ 
                                     &                    & Max & 16.07 & 16.64 & 36.47 \\ 
                            \cline{2-6}
                               & \multirow{3}{*}{Cervino} & Min & 13.49 & 11.99 & 52.66 \\ 
                                 &                        & Avg & 09.60 & 08.71 & 44.12 \\ 
                                 &                        & Max & 09.71 & 08.43 & 37.36 \\ 
     
 \hline
 \end{tabular}
 \caption{Sparbit percentage improvement of metrics over the second best algorithm.}
 \label{tab:improvement_values}
\end{table}
The first clear characteristic, as expected, is that the improvements under sequential mapping are
much more significant and consistent than the ones under cyclic mapping. This is outlined on
\Cref{fig:average_heatmaps} by a greater occurrence of darker tones for the plots of the latter on
both infrastructures. The same phenomenon can also be seen numerically on Table
\ref{tab:improvement_values}, as all the metrics present the smallest relative values on the cyclic
side. 
Focusing on the plots for sequential mapping, another interesting feature is that the highest
improvements appear generally on larger number of processes and data sizes. For Yahoo, above 128
processes and between 1KiB and 256KiB (287 tests), the mean (median) improvements are 60.64\%
(69.16\%). On Cervino, above 168 processes and on the same data size interval, the mean (median)
improvements are 43.86\% (45.65\%). The fact that on such cases the median is higher than the mean
indicates a distribution that is asymmetric to the left, and thus has more larger improvements than
lower ones. This difference is also higher on Yahoo, which supports the potential for growing
benefits on larger topologies. 

On a general picture the improvements are smaller but still very significant, as can be seen on
Table \ref{tab:improvement_values}. Under sequential mapping Yahoo and Cervino values are quite
similar, with means for the average time improvement respectively residing on about 35\% and 30\%,
highest near 80\% on both scenarios and average minimum as well as maximum times following similar
trends. Under cyclic mapping, the improvements for average time are reduced on both infrastructures,
achieving around 15\% on Yahoo and 10\% on Cervino. On this case, minimum and maximum metrics also
follow similar behaviours, but the highest obtained improvements are more variable across all
measurements ranging from 31\% to 52\%.

\section{Related work}
\label{sec:related_work}

The first set of related work is composed by the already presented and compared classic Allgather
algorithms (Section~\ref{sec:algorithms_problem}), namely Bruck \cite{bruck_efficient_1997},
Neighbor Exchange \cite{jing_chen_performance_2005}, Ring and Recursive Doubling.
Another form of improving the time of the Allgather operation, specially regarding communication
locality, is through rank reordering or remapping. On this approach the ranks assigned to a machine
are swapped by others in order to make the application's communication pattern better suited to the
topology and machine's hierarchy. One very direct form of this technique is the one presented by
\cite{alvarez-llorente_formal_2017}, which changes the mapping from sequential to cyclic and
vice-versa depending on the algorithm being executed. This approach shows great results for
converting adverse mapping scenarios to good ones, with only local transformation functions and very
little overhead. However, on the current form it requires gathering infrastructure information and
supposes only equal distribution of processes to machines, thus on highly heterogeneous environments
or with irregular mappings it would be unable to deliver great improvements. 

The works of \cite{mirsadeghi_topology-aware_2016} and \cite{kurnosov_dynamic_2016} have a more
Allgather focused approach, using its known communication pattern to create mappings more suited for
the algorithms. The first proposes fine-tuned heuristics for Ring, Recursive Doubling and Binomial
broadcast (a possible final component of an Allgather or Broadcast execution), with the experimental
results presenting improvements up to 78\%. It needs however, either initial sends or final memory
shifting to correct the blocks' positions, which adds overhead and even negatively degrades the
performance. The second work employs graph partitioning and linear optimization to fit the
communication pattern to the topology, presenting high potential improvements but requiring
infrastructure information and additional computation time, which grows for large numbers of
processes. 

The works of \cite{jeannot_near-optimal_2010} and \cite{li_topology-aware_2013} focus on mappings
for hierarchical tree networks and respectively propose the TreeMatch and Isomorphic Tree Mapping
algorithms. These approaches assume that the topology structure is a balanced tree and require the
profiling of applications in order to extract communication patterns. The information is then
employed to map the ranks via the introduced algorithms. Both proposals show improvements over
default forms of mapping, however require profiling information. In turn, the work of
\cite{jeannot_improving_2020} proposes online monitoring and rank remapping that provide
improvements and does not need prior executions, however it still requires active modification of
application code. Sparbit could be potentially coupled with these techniques, however its main
advantage in comparison is that it works out of the box, providing significant improvements on
communication time for theoretically any hierarchical network, and without need for topology
information, additional communication or computation.

\section{Conclusion}
\label{sec:conclusion}

On this paper we have highlighted the existing dichotomy of the employed Allgather algorithms for
MPI, in which the best theoretical ones have physical limitations, mainly regarding communication
locality, while the ones that present more local exchanges have a poor theoretical complexity. To
address this current situation, the Stripe Parallel Binomial Trees (Sparbit) algorithm has been
developed and presented, utilizing binomial trees to deliver the data faster, with optimal costs,
greater locality and no usage restrictions. 
Sparbit was empirically compared against the available algorithms on Open MPI 4.0.3, namely Bruck,
Recursive Doubling, Ring and Neighbor Exchange. The experiments were executed on two HPC
infrastructures with different topologies, network speeds and machine configurations to enrich the
data sources for the analysis. The results obtained indicate that Sparbit achieves the smallest
minimum, average and maximum times on a large fraction of test cases on the two infrastructures.
The sequential mapping is the one where Sparbit ameliorates more cases and does so more intensively,
however under cyclic a considerable amount of tests have also been improved.

The implementation employed here and geared towards the original MPI Allgather operation served as a
proof of concept for the potential benefits of the algorithm. Nonetheless, it can be easily
implemented to execute the Vector version of the call as well as the non-blocking forms, which we
plan to develop and benchmark as future work. Upon this expansion, Sparbit can be used on several
MPI collective operations.

\textbf{Acknowledgement:}
This study was financed in part by the Coordenação de Aperfeiçoamento de Pessoal de Nível Superior -
Brasil (CAPES) - Finance Code 001. It was also supported by FAPESC, UDESC, LabP2D, and University of
Neuch\^{a}tel.

\bibliographystyle{IEEEtran}
\bibliography{IEEEabrv,references}
\end{document}